\documentstyle[A4,12pt]{article}
\def\ds{\displaystyle}

\title{On direct numerical treatment of hypersingular integral
equations arising in mechanics and acoustics}
\author{Gerardo Iovane\\
D.I.I.M.A., University of Salerno, 84084 Fisciano (SA), Italy\\
e-mail: iovane@diima.unisa.it\\
\and I.K.\,Lifanov\\
The Zhukovsky Aeroforce Engineering Academy,\\
Leningradskii prospect 40, Moscow 125190, Russia\\
e-mail: lifanov\_ik@mail.ru\\
\and M.A. Sumbatyan\\
Faculty of Mechanics and Mathematics, Rostov\\
State University, Zorge 5, Rostov-on-Don 344090, Russia\\
e-mail: sumbat@math.rsu.ru}
\date{}

\begin{document}

\maketitle
\begin{abstract}

In this paper we present a treatment of hypersingular integral equations, which
have relevant applications in many problems of wave dynamics,
elasticity and fluid mechanics with mixed boundary conditions.
The main goal of the present work is the development of an efficient
direct numerical collocation method. The paper is completed with 
two examples taken from crack theory and acoustics: the study 
of a single crack in a linear isotropic elastic medium, and
diffraction of a plane acoustic wave by a thin rigid screen.
\end{abstract}

\section{Introduction}

Many problems of fluid mechanics, elasticity, and wave dynamics (acoustics)
with mixed boundary conditions can be reduced to hypersingular equations of
the following type
$$
\int\limits_{-1}^{1}\varphi \left( t\right) \left[ \frac{1}
{\left(x-t\right) ^{2}}%
+K_{0}\left( x,t\right) \right] dt=f\left( x\right) ,\qquad \qquad
\left| x\right| <1~,  
\eqno(1.1)
$$
where\ $\ K_0(x,t)$ \ is a regular ( in some sense) part of the kernel .
Direct numerical treatment of eq.(1.1) is not easy. Many authors applied
the approach, which may be currently considered as classical. If one defines
a solution bounded on the both endpoints \ $x=\pm 1$ \ ( that will
be shown below to vanish as 
$\sqrt{\left| x\pm 1\right| }$\ \ at \ $x\rightarrow \pm1$\ ), 
then one may search it in the form of the series
$$
\varphi \left( x\right) =\sqrt{1-x^{2}}\sum_{j=0}^{\infty }\varphi _{j}~%
U_{j}\left( x\right) ,\qquad \qquad \left| x\right| <1~,
\eqno(1.2)
$$
where ${U}_{j}\left( x\right) $ are the Chebyshev's polynomials of
the second kind [1]. Then, using the relation
$$
\begin{array}{c}
\ds\int\limits_{-1}^{1}\frac{\sqrt{1-t^{2}}}{\left( x-t\right) ^{2}}~{U}%
_{j}\left( t\right) dt=-\frac{d}{dx}~
\int\limits_{-1}^{1}\frac{\sqrt{1-t^{2}}}{x-t}%
~{U}_{j}\left( t\right) dt= \vspace*{3mm}\\
\ds=-\pi ~\frac{d}{dx}\left[ T_{j+1}\left( x\right) \right] =-\pi \left(
j+1\right) ~{U}_{j}~\left( x\right)
\end{array}
$$
( $T_{j}\left( x\right) $ is the Chebyshev's polynomial of the 
first kind), applying the scalar product of eq.(1.1) with the functions 
$\sqrt{1-x^{2~}}~{U}_{i}\left( x\right) $, one can reduce initial
integral equation (1.1) to the infinite linear algebraic system of the
second kind:
$$
-\left( i+1\right) \frac{\pi ^{2}}{2}\varphi _{i}+\sum_{j=0}^{\infty
}~k_{ij}~\varphi _{j}=f_{i}~,\qquad \qquad i=0,1,...,  
\eqno(1.3)
$$
$$
k_{ij}=\int\!\!\int_{-1}^{1}\sqrt{1-x^{2}}~\sqrt{1-t^{2}}~{U}_{i}\left(
x\right) ~{U}_{j}\left( t\right) ~K_{0}\left( x,t\right) dxdt~.
\eqno(1.4)
$$
We have used here the well known orthogonality relation [1]
$$
\int\limits_{-1}^{1}\sqrt{1-x^{2}}~{U}_{i}\left( x\right) ~{U}%
_{j}\left( x\right) =\frac{\pi }{2}\,\delta _{ij}~,\qquad \qquad
i,j=0,1,...,  
$$
where $\delta_{ij}$ is the Dirac delta.

The study of qualitative properties of the infinite system (1.3) is not the aim
of the present work, and we only note that, when the
regular kernel $K_0( x,t)$ is of a convolution type
[$K_{0}(x,t) =K_{0}(x-t)$], this approach is often applied in a different way.

We rewrite eq. (1.1) in equivalent form
$$
\int\limits_{-1}^{1}~K\left( x-t\right) ~\varphi \left( t\right) dt=f\left(
x\right)~,~ \left| x\right| <1~,  \quad
K\left( x\right) =\frac{1}{2\pi }~\int\limits_{-\infty }^{\infty }~L\left(
s\right) ~e^{-isx}~ds,  
\eqno(1.5)
$$
where $L\left( s\right) $ is the Fourier transform of $K\left( x\right) $. 
Let $L\left( s\right) $ be integrable for finite $s$, and for
$s\rightarrow \infty $ it has the following asymptotics
$$
L\left( s\right) =~A\left| s\right| +L_{0}\left( s\right) \quad ,\quad
L_{0}\left( s\right) =O\left( \frac{1}{s^{1+\delta }}\right) ,\qquad
\qquad (\delta >0)~,  
\eqno(1.6)
$$
then
$$
K\left( x\right) =-\frac{2A}{x^{2}}+K_{0}\left( x\right) ,\quad K_{0}\left(
x\right) =\frac{1}{2\pi }\int_{-\infty }^{\infty }L_{0}\left( s\right)
e^{-isx}ds~,  
$$
where $K_{0}\left( x\right) $ is at least continuous, and 
the generalized value of the integral [2] has been used, that is,
$$
\ds\int\limits_{-\infty }^\infty\left| s\right| e^{-isx}ds=2\lim_{\epsilon
\rightarrow +0}\int\limits_{0}^{\infty }e^{-\epsilon\,s}s\cos \left( sx\right)ds
=-\,\frac{2}{x^{2}}. 
\eqno(1.7)
$$

Equation (1.5) together with (1.6) is evidently equivalent to
eq.(1.1). If one substitutes the series expansion
(1.2) into eqs.(1.5) and then applies a scalar product with
functions $\sqrt{1-x^{2}}$ ${U}_{i}\left( x\right) $, one
finally arrives at the infinite linear algebraic system [1]
$$
\begin{array}{l}
\ds \sum_{j=0}^{\infty }l_{ij}\varphi _{j}=f_{i},\,(i=0,1,...),~
l_{ij}=\int\!\!\int\limits_{-1}^{1~}
\sqrt{\left(1-x^{2}\right)\left(
1-t^{2}\right)}{U}_{i}(x){U}_{j}(t)dxdt
\vspace*{3mm}\\
\ds\times\int\limits_{-\infty }^{\infty}L(s)e^{-is(x-t)}ds\sim
\int\limits_{-\infty }^{\infty }
\frac{J_{i+1}\left( s\right) J_{j+1}\left( s\right) L\left( s\right) ds}{s^{2}}.
\end{array}
\eqno(1.8)
$$
equivalent to system (1.3).
Note that the last integral is finite for all $i,j=0,1,...$,
since $J_{n}\left( s\right) \sim ~O\left( s^{n}\right) ,~s\rightarrow 0;~\quad
J_{n}\left( s\right) \sim O(1/\sqrt{s}) ,\quad
s\rightarrow \infty ,$
which, on taking into account the asymptotic property (1.6), is evident.

The integral equation (1.1) with a
hypersingular kernel can thus be reduced to an infinite system of
linear algebraic equations. This implies numerical treatment of double
integrals in (1.4) or single integrals over infinite interval of strongly
oscillating functions, like in (1.8). When solving integral
equations with more regular kernels, a powerful direct collocation technique,
in the framwork of the Boundary Element Method [3], can successfully be
applied. For the Cauchy-type integral equations, a direct collocation
method has been developed by Belotserkovsky and Lifanov [4, 5]. 
Hence, the main goal of the present paper is to develop an
efficient direct numerical collocation method.

\section{Some properties of hypersingular integrals}

First, one should clarify in which sense hypersingular integrals as given
by eq.(1.1) may be treated, since they do not exist either as improper
integrals of the first kind or as Cauchy-type singular integrals. At
least three different definitions of hypersingular integrals are known (see [6]):

1. The integral is a derivative of the Cauchy principal value
$$
\int\limits_{a}^{b}\frac{\varphi \left( t\right) }{\left( x-t\right)
^{2}}dt= -\frac{d}{dx}\int\limits_{\alpha }^{b}\frac{\varphi \left( t\right)}
{x-t}dt. 
\eqno(2.1)
$$

2. The integral is treated as a Hadamar principal value [4,5]
$$
\int\limits_{\alpha }^{b}\frac{\varphi \left(t\right)}{\left( x-t\right)^2}
dt=\lim_{\varepsilon \rightarrow +0}\left[ \left( \int\limits_{\alpha
}^{x-\varepsilon }+\int\limits_{x+\varepsilon }^{b}\right) \frac{\varphi
\left( t\right) dt}{\left( x-t\right)^2 }-\frac{2\varphi \left( x\right) }{%
\varepsilon }\right] .  
\eqno(2.2)
$$

3. The integral is a residue value, in the sense of generalized functions [2], that is
is an analytical continuation of the integral
$$
\int\limits_{\alpha }^{b}\left| x-t\right| ^{\alpha }\varphi \left(
t\right) dt 
\eqno(2.3)
$$
from domain, where it exists in the classical sense, to the value $%
\alpha =-2$ .

When $\varphi \left( x\right) \equiv 1$, the three different
approaches give the same result:

$$
\ds\int\limits_a^{b}\frac{dt}{\left( x-t\right) ^{2}}=-\,\frac{d}{dx}
\int\limits_{\alpha }^{b}\frac{dt}{x-t}=
\frac{d}{dx}\ln \left( \frac{b-x}{x-a}\right)=\frac{a-b}{\left( x-a\right)
\left( b-x\right) }~.
$$
$$
\int\limits_{a}^{b}\frac{dt}{\left( x-t\right) ^{2}}=
\lim_{\varepsilon \rightarrow +0}
\left[ \left( -\,\frac{1}{x-a}+\frac{1}{\varepsilon }+\frac{1}{%
\varepsilon }+\frac{1}{x-b}\right) -\,\frac{2}{\varepsilon }\right] =
\frac{a-b}{\left( x-a\right) \left( b-x\right) }~.
$$
$$
\ds\int\limits_{a}^{b}\left| x-t\right| ^{\alpha }dt=
\int\limits_{a}^{x}\left( x-t\right)^{\alpha }dt+
\int\limits_{x}^{b}\left( t-x\right) ^{\alpha }dt
=\frac{\left( x-a\right) ^{^{\alpha +1}}}{\alpha +1}+\frac{\left(
b-x\right) ^{^{\alpha +1}}}{\alpha +1}~,
\eqno(2.4)
$$
when ${Re}(\alpha )>-1.$ The analytical continuation of
(2.4) from the half-plane ${Re}\left( \alpha \right) >-1$ to the value $%
\alpha =-2$ gives
$$
\int\limits_{a}^{b}\frac{dt}{\left( x-t\right) ^{2}}=
\lim_{\alpha\to\,-2} \left[ \frac{\left(x-a\right) ^{\alpha+1}}{\alpha+1}+
\frac{\left( b-x\right) ^{\alpha +1}}{%
\alpha +1}\right] =
\frac{a-b}{\left( x-a\right) \left( b-x\right) }~.  
$$

All three definitions are equivalent to
each other, if the density $\varphi \left( x\right) $ is analytical (i.e.
differentiable) on the open interval $\left( a,b\right) $. Really, let
$\varphi(x)$ be analytic and ${Re}\left( \alpha \right) >1,$ then
$$
\begin{array}{l}
\ds -\frac{d}{dx}\int\limits_{a}^{b}\frac{\varphi \left( t\right)dt }{x-t}
=-\lim_{\varepsilon \to\,+0}\frac{d}{dx}\left(
\int\limits_{a}^{x-\varepsilon }+
\int\limits_{x+\varepsilon }^{b}\right) \frac{\varphi
\left( t\right) dt}{x-t}=
\lim_{\varepsilon \to\,+0}\left[ \left( \int\limits_{a}^{x-\varepsilon
}+\int\limits_{x+\varepsilon }^{b}\right) \frac{\varphi \left( t\right)dt} 
{\left(x-t\right) ^{2}}-\right.
\vspace*{3mm}\\
\ds \left. -\frac{2\varphi \left( x\right) }{\varepsilon }\right]~,\quad
\int\limits_{a}^{b}\left| x-t\right| ^{\alpha }\varphi \left( t\right)
dt=\lim_{\varepsilon \to +0}\left( \int\limits_{a}^{x-\varepsilon
}+\int\limits_{x+\varepsilon }^{b}\right) \left| x-t\right| ^{\alpha }\varphi
\left( t\right) dt=
\vspace*{3mm}\\
\ds=\lim_{\varepsilon\to +0}\left[ \int\limits_{a}^{x-\varepsilon }\left(
x-t\right) ^{\alpha }\varphi \left( t\right) dt+\int\limits_{x+\varepsilon
}^{b}\left( t-x\right) ^{\alpha }\varphi \left( t\right) dt\right] = 
\vspace*{3mm}\\
\ds=\frac{d}{dx}\lim_{\varepsilon\to +0}\left[ -\,\int\limits_{a}^{x-%
\varepsilon }\frac{\left( x-t\right) ^{\alpha +1}}{\alpha +1}\varphi \left(
t\right) dt+\int\limits_{x+\varepsilon }^{b}
\frac{\left( t-x\right) ^{\alpha +1}}{%
\alpha +1}\varphi \left( t\right) dt\right]~,
\end{array}
$$
so, by applying analytical continuation to the last relation, 
one can see (with the use of a standard 
$(\varepsilon,\delta)$ formalism) that the right-hand side results in (2.1).
Therefore, equivalence of 1 to 2 and 3 is evident. Let us prove that 
if $\varphi \left( x\right) \in C_{2}\left( a,b\right)$, 
then a finite value of the limit in expression (2.2) exists, and so
for $x\in \left( a,b\right) $ the integral $\int_{a}^{b}\varphi (t)/(x-t)^2\,dt$ 
is finite in any sense. Really, expression in the square brackets in eq.(2.2) is
$$
\left(\int\limits_a^{x-\varepsilon}+\int\limits_{x+\varepsilon}^b
\right) \frac{\varphi(t)-\varphi(x)-\varphi^{\prime }(x)(t-x)}
{(x-t)^2}dt
\ds+\varphi(x)\frac{a-b}{(x-a)(b-x)}+
\varphi ^{\prime }\left( x\right) \ln \frac{b-x}{x-a}~,
$$
which has a finite limit at $\varepsilon \to +0$.

\section{Integral equation with characteristic hyper-
singular kernel}

Consider hypersingular equation with the characteristic kernel
$$
\int\limits_a^b\frac{g\left( t\right) dt}{\left( x-t\right) ^{2}}=f'
\left( x\right) ,\qquad \qquad x\in \left( a,b\right) ,\quad f\left(
x\right) \in C_{2}\left( a,b\right)~.  
\eqno(3.1)
$$
Let us prove that a bounded solution of Eq.(3.1) is unique and is
given as follows
$$
g\left( x\right) =\frac{\sqrt{\left( x-a\right) \left( b-x\right) }}{\pi ^{2}%
}\int\limits_{a}^{b}\frac{f\left( t\right) dt}{\sqrt{\left( t-a\right) \left(
b-t\right) }\,\left( x-t\right) }~.
$$
Really, according to the definition (2.1), equation
(3.1) is equivalent to
$$
\frac{d}{dx}\int\limits_{a}^{b}\frac{g\left( t\right) dt}{\left( x-t\right) }%
=-f^{\prime }\left( x\right) ,\quad \sim \quad
\int\limits_{a}^{b}\frac{g(t) dt}{x-t}=-f\left( x\right) +C,
\qquad x\in \left( a,b\right)~,  
$$
where $C$ is an arbitrary constant. Now an inversion formula for the Cauchy 
characteristic integral operator [7] determines the bounded solution as
$$
g\left( x\right) =\frac{\sqrt{\left( x-a\right) \left( b-x\right) }}{\pi ^{2}%
}\int\limits_{a}^{b}\frac{f\left( t\right)}{\sqrt{\left( t-a\right) \left(
b-t\right) }\,\left( x-t\right) }dt,\quad x\in (a,b)~,
\eqno(3.2)
$$
and the constant $C$ as
$$
C =\frac{1}{\pi ^{2}}
\int\limits_{a}^{b}\frac{f\left( t\right)}{\sqrt{\left( t-a\right) \left(
b-t\right) }}dt~.
$$
Thus, as indicated in the Introduction, any bounded solution of Eq.(3.1) 
vanishes at $x\rightarrow a,b$.

To construct a direct collocation technique to
solve eq.(3.1) for arbitrary right-hand side,
we divide the interval $\left( a,b\right) $ into $n$ small equal subintervals
of the length $h=\left( b-a\right) /n,$ by the nodes 
$a=t_{0},\,t_{1},\,t_{2},...,t_{n-1,\,}t_{n}=b~,~
t_{j}=a+jh,\quad j=0,1...,n.$ The central points of each
sub-interval $\left( t_{i-1},t_{i}\right) $ are denoted by $x_{i}$, thus $%
x_{i}=a+\left( i-1/2\right) h\,,~ i=1,...,n.$

If we try to arrange an approximation of integral in (3.1) by using finite
sum, like in regular cases, then for $x=x_{i}$ we have
$$
\begin{array}{c}
\ds\int_{a}^{b}\frac{g\left( t\right) dt}{\left( x_{i}-t\right) ^{2}}%
\approx \sum_{j=0}^{n}\,g\left( t_{j}\right)\, 
\int\limits_{t_{j-1}}^{t_{j}}\frac{%
dt}{\left( x_{i}-t\right) ^{2}}=g\left( t_{i}\right) \int_{-h/2}^{h/2}\frac{dt}{t^{2}}+
\vspace*{3mm}\\
\ds+\sum_{j\neq i}g(t_{j}) \left( \frac{1}{x_{i}-t_{j}}-\frac{1}{%
x_{i}-t_{j-1}}\right)=
\sum_{j=1}^{n}g(t_{j}) \left( \frac{1}{x_{i}-t_{j}}-\frac{1}{%
x_{i}-t_{j-1}}\right)~,
\end{array}
\eqno(3.3)
$$
where a value of the above hypersingular integral of the function 
$1/t^{2}$ has been used. So, we try to approximate eq.(3.1)
by the linear algebraic system 
$$
\sum_{j=1}^{n}g\left( t_{j}\right) \left( \frac{1}{x_{i}-t_{j}}-\frac{1}{%
x_{i}-t_{j-1}}\right) =f^{\prime }(x_i),\qquad \qquad i=1,...,n.  
\eqno(3.4)
$$

Further considerations are similar to those in [4,5]. Let us prove that, by 
assuming $x=x_l \in \left( a,b\right) $ is fixed, the
difference between solution $g( x_l)$ of the system (3.4)
and analytical solution (3.2) tends to zero, when $n\rightarrow \infty $.
Obviously, the system (3.4) can be rewritten as
$$
\sum_{j=1}^{n}g\left( t_{j}\right) \left( \frac{1}{x_{j}-t_{i}}-\frac{1}{%
x_{j}-t_{i-1}}\right) =f'\left( x_{i}\right) ,\qquad \qquad
i=1,...,n~, 
\eqno(3.5)
$$
and its principal determnant is
$$
\Delta =\left| 
\begin{array}{ccc}
\ds\left( \frac{1}{x_{1}-t_{1}}-\frac{1}{x_{1}-t_{0}}\right) & ... &\ds\left( 
\frac{1}{x_{n}-t_{1}}-\frac{1}{x_{n}-t_{0}}\right) \\ 
... & ... & ... \\ 
\ds\left( \frac{1}{x_{1}-t_{n}}-\frac{1}{x_{1}-t_{n-1}}\right) & ... & 
\ds\left( \frac{1}{x_{n}-t_{n}}-\frac{1}{x_{n}-t_{n-1}}\right)
\end{array}
\right| .  
\eqno(3.6)
$$

Let us rewrite the $i-th$ row $\left( i=2,...,n\right) $ of $\Delta $ as
a sum of all other rows, from $k=1$ to $k=i$:
$$
\begin{array}{l}
\ds \Delta =\left| 
\begin{array}{ccc}
\ds\frac{t_{1}-t_{0}}{\left( x_{1}-t_{1}\right) \left( x_{1}-t_{0}\right) } & 
... & 
\ds\frac{t_{1}-t_{0}}{\left( x_{n}-t_{1}\right) \left( x_{n}-t_{0}\right) 
} \\ 
... & ... & ... \\ 
\ds\frac{t_{n}-t_{0}}{\left( x_{1}-t_{n}\right) \left( x_{1}-t_{0}\right) } & 
... & 
\ds\frac{t_{n}-t_{0}}{\left( x_{n}-t_{n}\right) \left( x_{n}-t_{0}\right) 
}
\end{array}
\right| =
\ds\frac{\prod\limits_{j=1}^n\left( t_{j}-t_{0}\right) }
{\prod\limits_{i=1}^n\left( x_{i}-t_{0}\right) }\times
\vspace*{4mm}\\
\ds \times\left| 
\begin{array}{rrr}
\ds\frac{1}{x_{1}-t_{1}} & ... & \ds\frac{1}{x_{n}-t_{1}} \\ 
... & ... & ... \\ 
\ds\frac{1}{x_{1}-t_{n}} & ... & \ds\frac{1}{x_{n}-t_{n}}
\end{array}
\right| =
\ds \frac{\prod\limits_j\left( t_{j}-t_{0}\right) }
{\prod\limits_i\left( x_{i}-t_{0}\right) }~
\frac{\prod\limits_{q\,<}\!\prod\limits_{p}\left(t_{q}-t_{p}\right)
\prod\limits_{q\,<}\!\prod\limits_{p}\left(x_{p}-x_{q}\right) }
{\prod\limits_{q\,,}\!\prod\limits_{p}\left( x_{q}-t_{p}\right) },
\end{array}
\eqno(3.7)
$$
where the lower limit in all products is $1$ and the upper is $n$. A 
known value of the last determinant has been taken from [4,5].

According to the Cramer's rule, one needs to calculate the determinant $%
\Delta _{l}$ where a column with the right-hand side (3.5)
is substituted instead of the $l-th$ column of $\Delta $ (3.6):
$$
\begin{array}{c}
\ds\Delta _{l}=\left| 
\begin{array}{ccccc}
\ds\left( \frac{1}{x_{1}-t_{1}}-\frac{1}{x_{1}-t_{0}}\right) & ... & 
\ds f^{\prime }\left( x_{1}\right) & ... & 
\ds\left( \frac{1}{x_{n}-t_{1}}-\frac{1}{%
x_{n}-t_{0}}\right) \\ 
\ds\left( \frac{1}{x_{1}-t_{2}}-\frac{1}{x_{1}-t_{1}}\right) & ... & 
\ds f^{\prime
}\left( x_{2}\right) & ... & 
\ds\left( \frac{1}{x_{n}-t_{2}}-\frac{1}{%
x_{n}-t_{1}}\right) \\ 
... & ... & ... & ... & ... \\ 
\ds\left( \frac{1}{x_{1}-t_{n}}-\frac{1}{x_{1}-t_{n-1}}\right) & ... & 
\ds f^{\prime }\left( x_{n}\right) & ... & 
\ds\left( \frac{1}{x_{n}-t_{n}}-\frac{1}{%
x_{n}-t_{n-1}}\right)
\end{array}
\right|=
\vspace*{4mm}\\
\ds=\left| 
\begin{array}{ccccc}
\ds\frac{t_{1}-t_{0}}{\left( x_{1}-t_{1}\right) \left( x_{1}-t_{0}\right) } & 
... & 
\ds f^{\prime }\left( x_{1}\right) & ... & 
\ds\frac{t_{1}-t_{0}}{\left(
x_{n}-t_{1}\right) \left( x_{n}-t_{0}\right) } \\ 
\ds\frac{t_{2}-t_{0}}{\left( x_{1}-t_{2}\right) \left( x_{1}-t_{0}\right) } & 
... & 
\ds f' \left( x_{1}\right) +f^{\prime }\left( x_{2}\right) & ... & 
\ds\frac{t_{2}-t_{0}}{\left( x_{n}-t_{2}\right) \left( x_{n}-t_{0}\right)}\\ 
... & ... & ... & ... & ... \\ 
\ds\frac{t_{n}-t_{0}}{\left( x_{1}-t_{n}\right) \left( x_{1}-t_{0}\right) } & 
... & 
\ds\sum\limits_{k=1}^{n}f^{\prime }\left( x_{k}\right) & ...& 
\ds\frac{t_{n}-t_{0}}{\left( x_{n}-t_{n}\right) \left( x_{n}-t_{0}\right) }
\end{array}
\right| ,  
\end{array}
$$
where we have used the same summation of rows as in the case of $\Delta $.

The last determinant may be calculated arranging expansion by elements of
the $l-th$ column as follows (see also [4,5])
$$
\Delta _{l}=\frac{\prod\limits_j\left( t_{j}-t_{0}\right) }
{\prod\limits_{i\neq l}\left( x_{i}-t_{0}\right) }~
\sum\limits_{m=1}^{n}
\frac{\sum\limits_{k=1}^m f^{\prime }\left( x_{k}\right) }
{t_{m}-t_{0}}\left(
-1\right) ^{m+l}\frac{\prod\limits_{q<p;}\,\prod\limits_{q,p\neq m}
\left(t_{q}-t_{p}\right) 
\prod\limits_{q<p;}\,\prod\limits_{q,p\neq l}
\left(x_{p}-x_{q}\right) }
{\prod\limits_{q\neq l}\prod\limits_{p\neq m}
\left(x_{q}-t_{p}\right) }~,  
$$
so
$$
\ds g(x_{l})=\frac{\Delta _{l}}{\Delta }=(x_{l}-t_{0})\sum_{m=1}^{n}
\frac{\sum\limits_{k=1}^m f^{\prime }(x_{k})\quad
\prod\limits_{p}\left( x_{l}-t_{p}\right) ~
\prod\limits_{q}\left( x_{q}-t_{m}\right) ~}
{(t_{m}-t_{0})(x_{l}-t_{m})\prod\limits_{p\neq m}\left( t_{m}-t_{p}\right) ~
\prod\limits_{q\neq l}\left( x_{q}-x_{l}\right) }~.
\eqno(3.8)
$$

The last expression at $h\rightarrow 0\sim n\rightarrow \infty 
$, under the condition $x_{l}\in \left( a,b\right) $ is fixed, can be estimated
as follows:
$$
\begin{array}{c}
\ds\frac{\prod\limits_{p=1}^n\left( x_{l}-t_{p}\right) }
{\prod\limits_{q\neq l}\left( x_{q}-x_{l}\right) }
=\left( x_{l}-t_{l}\right) 
\frac{\prod\limits_{p=1}^{l-1}\left( x_{l}-t_{p}\right) }
{\prod\limits_{q=1}^{l-1}\left(x_{q}-x_{l}\right) }\times 
\frac{\prod\limits_{p=l+1}^n\left( x_{l}-t_{p}\right) }
{\prod\limits_{g=l+1}^{n}\left( x_{q}-x_{l}\right) }=
\vspace*{4mm}\\
\ds=\frac{(-1)^n h}{2}
\prod\limits_{p=1}^{l-1}\left( 1-\frac{1%
}{2p}\right) .\prod\limits_{p=1}^{l-1}\left( 1+\frac{1}{2p}\right)
\ds\sim \left( -1\right) ^{n}\frac{h\sqrt{b-x_{l}}}{\pi\sqrt{x_{l}-a}}~,~n\to\infty~,
\end{array}
\eqno(3.9)
$$
if $x$ belongs to the open interval $\left( a,b\right) \quad \left(
i.e.\quad l\neq n,\quad l\neq 1\right) .$ Here we have used the asymptotic estimate [4,5]
$$
\prod_{m=1}^{n}\left( 1+\frac{\beta }{m}\right) =\frac{n^{\beta }}{\Gamma
\left( 1+\beta \right) }+O\left( n^{\beta -1}\right) ,\quad n\rightarrow
\infty~,  
$$

Further, by analogy
$$
\frac{\prod\limits_{q}\left( x_{q}-t_{m}\right) }
{\prod\limits_{p\neq m}\left( t_{m}-t_{p}\right) } =
\left( -1\right) ^n\left(t_{m}-x_{m}\right) 
\prod\limits_{q=1}^{m-1}\left( 1+\frac{1}{2q}\right)
\prod\limits_{q=1}^{n-m}\left( 1-\frac{1}{2q}\right)
\sim \left( -1\right) ^{n}\frac{h\sqrt{t_{m}-a}}{\pi\sqrt{b-t_{m}}}.
$$

Other terms in eq. (3.8) can be simplified as follows ($h\rightarrow 0$):
$$
(x_{l}-t_{0})\rightarrow (x_{l}-a),~
(t_{m}-t_{0})\rightarrow (t_{m}-a),~ 
h\sum\limits_{k=1}^{m}f^{\prime }(t_{k}) \rightarrow
\int\limits_{a}^{t_{m}}f^{\prime }(t)dt=f(t_{m}),
$$
hence expression (3.8) with $h\rightarrow 0$ tends to
$$
\begin{array}{c}
\ds g\left( x_{l}\right) \sim \frac{h}{\pi ^{2}}\sqrt{\left( x_{l}-a\right)
\left( b-x_{l}\right) }\,\sum\limits_{m=1}^{n}
\frac{f\left( t_{m}\right) }
{\sqrt{\left( t_{m-a}\right) \left( b-t_{m}\right) }
~\left( x_{l-}t_{m}\right) } \sim
\vspace*{3mm} \\
\ds\sim \frac{\sqrt{\left( x_{l}-a\right) \left( b-x_{l}\right) }}{\pi ^{2}}
\,\int\limits_{a}^{b}\frac{f\left( t\right) dt}{\sqrt{\left( t-a\right) \left(
b-t\right) }~\left( x_{l}-t\right) }~.
\end{array}
$$

An example on application of the proposed numerical method 
is shown in Fig.1, where $(a,b)=(-1,1)$.

\section{Full hypersingular equation}

Consider the full equation
$$
\int\limits_{a}^{b}
\left[ \frac{1}{\left( x-t\right) ^{2}}+K_{0}\left( x,t\right) %
\right] g\left( t\right) dt=f^{\prime }\left( x\right) ,\quad 
x\in \left( a,b\right)~ ,  
\eqno(4.1a)
$$
where
$$
K_{0}\left( x,t\right) =\frac{\partial K_{1}\left( x,t\right) }{\partial x}.
\eqno(4.1b)
$$
Its bounded solution can be constructed by applying inversion of
the characteristic part, that reduces eq.(4.1a) to a second-kind Fredholm
integral equation
$$
g\left( x\right) +\int\limits_{a}^{b}N_{1}\left( x,t\right) g\left( t\right)
dt=f_{1}\left( x\right) ,\quad x\in \left( a,b\right)~ , 
\eqno(4.2)
$$
where
$$
N_{1}\left( x,t\right) =\frac{\sqrt{\left( x-a\right) \left( b-x\right) }}{%
\pi ^{2}}\int\limits_{a}^{b}\frac{K_{1}\left( \tau ,t\right) d\tau }
{\sqrt{\left(\tau -a\right) \left( b-\tau \right) }~
\left( x-\tau \right) }~,  
$$
$$
f_{1}\left( x\right) =\frac{\sqrt{\left( x-a\right) \left( b-x\right) }}{\pi
^{2}}\int\limits_{a}^{b}
\frac{f\left( \tau \right) d\tau}{\sqrt{\left( \tau -a\right)
\left( b-\tau \right) }~\left( x-\tau \right) }~.  
$$

Let us prove that, if $f\left( x\right) \in C_{1}$\bigskip 
$(a,b);~K_{1}\left( x,t\right) \in C_{1}\left[ \left(
a,b\right) \times \left( a,b\right) \right]$, then for any $x\in \left(
a,b\right) $ the difference between solution $g\left( x\right) $ of the
linear algebraic system
$$
\sum\limits_{j=1}^{n}\left[ \frac{1}{x_{i}-t_{j}}-
\frac{1}{x_{i}-t_{j-1}}+
hK_0\left( x_{i},t_{j}\right) \right]\, g\left(
t_{j}\right) =f^{\prime }\left( x_{i}\right) ,~i=1,...,n
$$
and the bounded solution of eq.(4.1) tends to zero when $h\rightarrow
0 $ (i.e. $n\rightarrow \infty $). Really, if one
transfers the terms, related to the regular kernel, to the right-hand side
and solves so written linear algebraic system with the characteristic matrix 
$1/\left( x_{i}-t_{j}\right) -1/\left( x_{i}-t_{j-1}\right) $, then one
arrives at a finite-difference approximation of the eq.(4.2). The
proof is finally completed, if one applies classical results on numerical
solution of the second-kind Fredholm integral equation.

An example for $K_{0}\left( x,t\right) =A\left( x-t\right),~
f^{\prime }\left( x\right) =-\pi \sim f\left( x\right)=-\pi x$ is shown
in Fig.2, where numerical solution is compared with the exact one
$g\left( x\right) =8\sqrt{1-x^{2}}\left( 4+Ax\right)/(32+A^{2})$
in the case $A=3, ~\left( a,b\right) =\left( -1,1\right)$.

It should also be noted that some interesting results on the theory and
numerical treatment of hypersingular integral equations have been recently
obtained in [8-10].

\section{Some examples from mechanics and acoustics}

\subsection{\protect\bigskip\ Single crack in linear isotropic elastic medium}

Consider the two-dimensional in-plane problem. Then the
displacement field is ${\bf\bar{u}}=\left( u_{x,},u_{y},0\right)$, where 
$u_{x}=u_{x}\left( x,y\right) $, $u_{y}=u_{y}\left( x,y\right)$ can be
represented in the Papkovich-Neuber form
$$
u_{x}=\frac{\partial \varphi }{\partial x}+y\,
\frac{\partial \psi }{\partial x},\quad 
u_{y}=\frac{\partial \varphi }{\partial y}+
y\,\frac{\partial \psi }{\partial y}-\chi\,\psi~,  \quad
\chi=3-4\nu~,
$$
where the potentials $\varphi $ and $\psi $ are harmonic
$$
\Delta \varphi =0,\qquad \Delta \psi =0~,
\eqno(5.1)
$$
and $\nu=\lambda/2(\lambda+\mu)$ is the Poisson ratio ($\lambda$ and
$\mu $ are elastic constants).

Let a constant normal load $\sigma _{yy}=\sigma _{0}$ $\ \left( \tau _{xy}=0\right)$ 
be applied symmetrically to internal faces of the
straight-line crack $y=0, ~x\in \left(-a,a\right)$.
Then, due to a symmetry, the problem can be reduced to the upper
half-plane $y\ge 0$ with the boundary conditions on the boundary line $y=0$
$$
\tau _{xy}=0,~ -\infty <x<\infty~;\quad
u_{y}=0, ~ \left| x\right| >a~;\quad
\sigma _{yy}=-\sigma _{0}, ~ \left| x\right| <a~.
\eqno(5.2)
$$

Solution of eqs.(5.1) can be constructed with the use of Fourier
transform (all capital letters are Fourier transforms of corresponding functions):
$$
\Phi \left( s,y\right) =A\left( s\right) e^{-\left| s\right| y},\quad \Psi
\left( s,y\right) =B\left( s\right) e^{-\left| s\right| y}~,
$$
where the well known properties
$$
f\left( x\right) \Longrightarrow F\left( s\right) ,~
f^{\prime \prime }\left( x\right) \Longrightarrow -s^{2}F\left( s\right),~
{\rm so}~ \Delta\left( \varphi ,\psi \right) \Longrightarrow \frac{d^{2}}{dy^{2}}
\left(\Phi ,\Psi \right) -s^{2}\left( \Phi ,\Psi \right)
$$
have been used.

The two constants $A(s)$ and $B(s)$ can be determined from the boundary
conditions (5.2) using
$$
\tau _{xy}=\mu \left[2\frac{\partial ^{2}\varphi }{\partial
x\partial y}+(1-\chi)\,\frac{\partial \psi}{\partial x}\right]~, \quad
\sigma _{yy}=2\mu\frac{\partial ^{2}\varphi }{\partial y^2}+
(1-\chi)(\lambda+2\mu)\frac{\partial\psi}{\partial y}~.  
\eqno(5.3)
$$

It follows from (5.2)$_1$ that
$$
2(-is)\,(-|s|)\, A(s)+(1-\chi)\,(-is)\,B(s)=0
\Longrightarrow ~A(s) =\frac{1-\chi}{2|s|}\,B(s)~.
\eqno(5.4)
$$

Introducing a new function
$$
g(x) = 
\left\{
\begin{array}{ll}
u_{y}~,\qquad\qquad |x| < a \\ 
u_{y}=0~, \qquad |x| >a
\end{array}
\right.~,\quad {U}_{y}\left( s,0\right) =\int\limits_{-a}^{a}g\left( t\right) e^{ist}dt 
\eqno(5.5)
$$
using eq.(5.4), we obtain the constants $A\left(s\right)$
and $B\left( s\right) $ in terms of $g\left( x\right)$
$$
A\left( s\right) =\frac{\chi-1}{|s|\,(\chi+1)}\,
\int\limits_{-a}^{a}g\left( t\right)
e^{ist}dt~,\qquad B\left( s\right) =-\,\frac{2}{\chi+1}\,
\int\limits_{-a}^{a}g\left( t\right) e^{ist}dt~.  
$$

>From (5.3) one obtains the Fourier transform of the normal stress $\sigma_{yy}$:
$$
\Sigma_{yy}\left( s,0\right) =2\mu s^{2}\,A(s)+
(1-\chi)(\lambda+2\mu)\,(-|s|)\,B(s)
=-\frac{\mu}{1-\nu}\,|s|\,\int\limits_{-a}^{a}g\left( t\right) e^{ist}dt~,  
$$
and the boundary condition (5.2)$_3$ leads thus to the integral equation
($\left| x\right| <a$)
$$
\int\limits_{-a}^{a}g\left( t\right) K\left( t-x\right) dt=
\frac{2\pi(1-\nu)}{\mu}\,\sigma_{0}~,\quad
K\left( x\right) =\int\limits_{-\infty }^{\infty }\left| s\right| 
e^{-isx}ds=-\frac{2}{x^{2}}~.
\eqno(5.6)
$$

Equation (5.6) is the characteristic hypersingular integral
equation
$$
\int\limits_{-a}^{a}\frac{g\left( t\right) }{\left( x-t\right) ^{2}}\,dt=
-\,\frac{\pi(1-\nu)}{\mu }\,\sigma _{0}~,\quad ~\left| x\right| <a  
\eqno(5.7)
$$
with respect to the function $g\left( x\right) $ which is, due to eq.(5.5),
the relative opening of the crack's faces. Solution of eq.(5.7) is (see Section 3)
$$
g\left( x\right) =u_{y}\left( x,0\right) =\frac{\sigma _{0}}{\mu }
(1-\nu)\sqrt{a^{2}-x^{2}~,}\quad ~\left| x\right| <a~,  
\eqno(5.8)
$$
which can also be obtained by the proposed direct numerical collocation
method (see Fig.1). Note that (5.8) coincides with a classical solution of the
considered problem.

\subsection{Diffraction by a thin rigid screen}

Acoustically hard thin screen is placed horizontally on the
interval $-a<x<a,~y=0$. A plane acoustic wave is normally incident on the screen:
$p^{inc}=e^{iky}~,\quad k=\omega/c$~,  
where $\omega $ is the angular wave freguency, $c$ is the wave speed. The
total pressure is then a sum of the incident and the scattered wave fields
$p\left( x,y\right) =p^{inc}+p^{sc}$~,
where all three components of the wave field satisfy Helmholtz equation
$$
\Delta p+k^{2}p=0~.
\eqno(5.9)
$$

By analogy to the previous problem, 
the Fourier images in the lower $\left( x<0\right) $ and upper $\left(
x>0\right) $ half-plane are, respectively ($\gamma \left( s\right) =\sqrt{s^{2}-k^{2}}$):
$$
P_{-}^{sc}\left( s,y\right) =A\left( s\right) e^{\gamma \left( s\right)
y\quad },~ y<0~;\quad
P_{+}^{sc}\left( s,y\right) =B\left( s\right) ~e^{-\gamma \left( s\right)
y}\quad ,~ y>0~.
\eqno(5.10)
$$
The coefficients $A$ and $B$ are defined from boundary
conditions at $y=0$
$$
\frac{\partial p_{+}\left( x,0\right) }{\partial y}=\frac{\partial p_{-}(x,0)
}{\partial y}=0~,~
\frac{\partial p_{+}^{sc}\left(
x,0\right) }{\partial y}=\frac{\partial p_{-}^{sc}\left( x,0\right) }{%
\partial y}=-ik~,~ \left| x\right| <a~,
\eqno(5.11)
$$
$$
\frac{\partial p_{+}\left( x,0\right) }{\partial y}=\frac{\partial p_{-}(x,0)%
}{\partial y}~,\qquad
 \frac{\partial p_{+}^{sc}\left( x,0\right) }{\partial y}=%
\frac{\partial p_{-}^{sc}\left( x,0\right) }{\partial y}~,\quad \left|
x\right| >a  
\eqno(5.12)
$$
$$
p_{+}\left( x,0\right) =p_{-}(x,0)~,\qquad p_{+}^{sc}\left( x,0\right)
=p_{-}^{sc}\left( x,0\right)~ ,\quad \left| x\right| >a  ~.
\eqno(5.13)
$$

The conditions (5.11), (5.12) imply that
$\partial p_{+}^{sc}/\partial y=\partial p_{-}^{sc}/\partial y$
for $y=0$ and all $ -\infty<x<\infty$, thus
$\gamma A=-\gamma B\Longrightarrow B\left( s\right) =-A\left( s\right)$.
Now it is obvious from (5.10) that $p_{-}^{sc}\left( x,0\right) =-p_{+}^{sc}\left(
x,0\right) $. It thus implies, together with (5.13), that
$p_{-}^{sc}\left( x,0\right) =-p_{+}^{sc}\left( x,0\right) =0,~\left|x\right| >a$.

Introducing a new function $g\left( x\right)$
$$
p^{sc}\left( x,0\right) = 
\left\{
\begin{array}{rr}
0~,\qquad \left| x\right| >a \\ 
g\left( x\right)~ ,\quad \left| x\right| <a~,
\end{array}
\right.  
$$
it follows from (5.10) that
$$
A\left( s\right) =\int\limits_{-a}^{a}g\left( t\right) e^{ist}dt~,  
$$
and the boundary condition (5.11) leads to the integral
equation involving function $g\left( x\right) $
$$
\int\limits_{-a}^{a}g\left( t\right) K\left( x-t\right) dt=-ik,\quad \left|
x\right| <a~,\qquad
K\left( x\right) =\frac{1}{2\pi }\int\limits_{-\infty }^{\infty }
\gamma \left(s\right) e^{-isx}ds~,
\eqno(5.14)
$$
where the kernel
$$
\begin{array}{c}
\ds K\left( x\right) =\frac{1}{2\pi }\int\limits_{-\infty }^{\infty }
\left| s\right|e^{-isx}ds+
\frac{1}{2\pi }\int\limits_{-\infty }^{\infty }
\left( \sqrt{s^{2}-k^{2}}-\left| s\right| \right) e^{-isx}ds=  
\vspace{3mm}\\
\ds=-\frac{1}{\pi x^{2}}+K_{0}\left( x\right)~,\quad
K_{0}\left( x\right)=
\frac{1}{\pi }\int\limits_{0}^{\infty }
\left( \sqrt{s^{2}-k^{2}}-s\right) \cos \left( sx\right) ds~,
\end{array}
\eqno(5.15)
$$
and the regular kernel $K_{0}\left( x\right) $ is differentiable for all $%
x\neq 0$, with a weak singularity in origin: $K_{0}\left( x\right) =O\left( \ln \left|
x\right|\right)~,~x\to 0$, that permits application of the proposed numerical method.
An example for some values of $k$ is shown in Fig.3.

\section{Conclusions}

\quad 1. The proposed method is an efficient alternative to a standard
reduction to infinite systems 
of linear algebraic equations. This is based on direct numerical
treatment of hypersingular integrals, and reduces the problem to
a finite system of linear algebraic equations. Its principal merit
is that there is no need in numerical
computations when calculating elements of respective matrix.

2. In some cases the proposed method permits explicit analytical
solution of respective hypersingular integral equation.
Otherwise, the proposed method provides an efficient numerical
treatment, and convergence of the algorithm has been proved when
the step of the mesh tends to zero.

3. The considered examples show a good precision of the method, since
with a few decades of nodes on the mesh the obtained numerical results
almost coincide with respective analytical solution, in the cases when
the latter is known.

{\bf Acknowledgements}

The paper has been supported by Italian Ministry for Instructions, University
and Research (M.I.U.R.) through its national and local projects.

\newpage

\begin{center}
{\Large Legends to Figures}
\end{center}

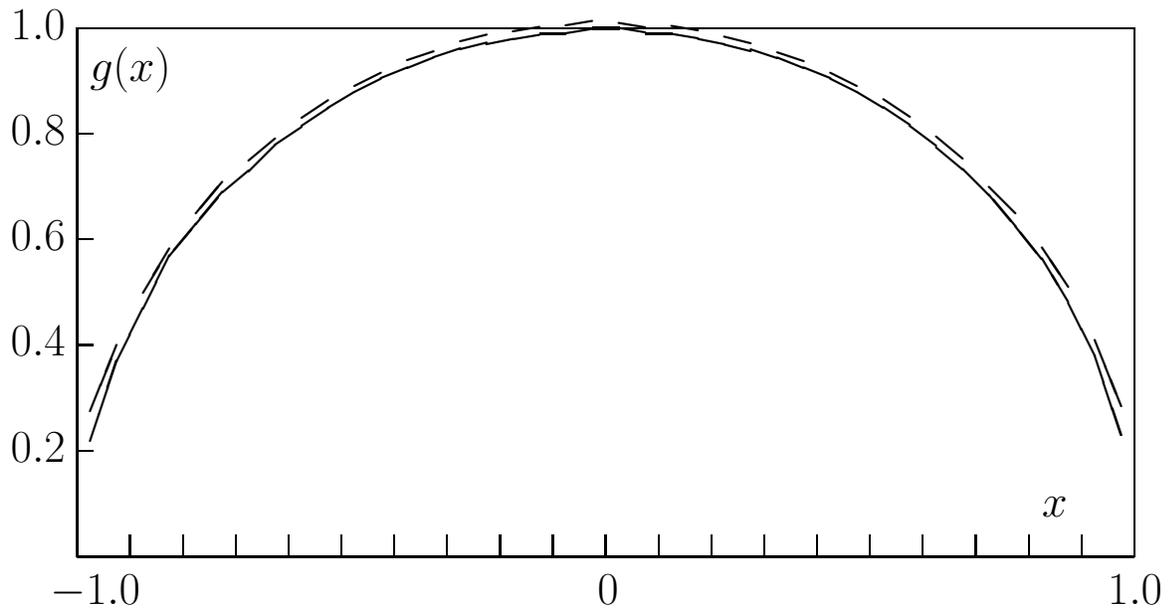
Figure 1. Comparison between exact and numerical solutions of the
characteristic equation with $f^{\prime}(x)=-\pi \quad \sim \quad
f(x)=-\pi\, x$: --- exact solution $g(x)=\sqrt{1-x^2}$,\quad - - - numerical
solution; $n=40$.

\vspace*{2cm} Figure 2. Comparison between exact and numerical solutions of
the full equation with $f^{\prime}(x)=-\pi \quad \sim \quad f(x)=-\pi\, x$%
,\quad $K_0(x,t)=A(x-t),\quad A=3$: --- exact solution,\quad - - - numerical
solution; $n=40$.

\vspace*{2cm} Figure 3. Comparison between exact and numerical solutions of
the equation (5.14)-(5.15),\quad ($a=1$): 1 - $k=0.5$,\quad
2 - $k=1.5$,\quad 3 - $k=2.5$.\quad The factor $-\pi ik$ is omitted.

\newpage \vspace*{2cm}
\qquad\qquad {\Large 
\begin{picture}(300,200)

\thinlines
\put(-50,0){\line(1,0){400}}
\put(-50,200){\line(1,0){400}}
\put(-50,0){\line(0,1){200}}
\put(350,0){\line(0,1){200}}
\put(-30,0){\line(0,1){8}}
\put(-10,0){\line(0,1){8}}
\put(10,0){\line(0,1){8}}
\put(30,0){\line(0,1){8}}
\put(50,0){\line(0,1){8}}
\put(70,0){\line(0,1){8}}
\put(90,0){\line(0,1){8}}
\put(110,0){\line(0,1){8}}
\put(130,0){\line(0,1){8}}
\put(150,0){\line(0,1){8}}
\put(170,0){\line(0,1){8}}
\put(190,0){\line(0,1){8}}
\put(210,0){\line(0,1){8}}
\put(230,0){\line(0,1){8}}
\put(250,0){\line(0,1){8}}
\put(270,0){\line(0,1){8}}
\put(290,0){\line(0,1){8}}
\put(310,0){\line(0,1){8}}
\put(330,0){\line(0,1){8}}

\put(-50,40){\line(1,0){6}}
\put(-75,36){$0.2$}
\put(-50,80){\line(1,0){6}}
\put(-75,76){$0.4$}
\put(-50,120){\line(1,0){6}}
\put(-75,116){$0.6$}
\put(-50,160){\line(1,0){6}}
\put(-75,156){$0.8$}
\put(-75,196){$1.0$}
\put(-45,180){$g(x)$}

\put(315,15){$x$}
\put(147,-18){$0$}
\put(-60,-18){$-1.0$}
\put(340,-18){$1.0$}

\thicklines
\put(-45, 44){\line(  1,  3){ 10}}
\put(-35, 74){\line(  1,  2){ 10}}
\put(-25, 94){\line(  1,  2){ 10}}
\put(-15,114){\line(  5,  6){ 10}}
\put( -5,126){\line(  5,  6){ 10}}
\put(  5,138){\line(  6,  5){ 10}}
\put( 15,146){\line(  1,  1){ 10}}
\put( 25,156){\line(  3,  2){ 10}}
\put( 35,163){\line(  3,  2){ 10}}
\put( 45,170){\line(  5,  3){ 10}}
\put( 55,176){\line(  2,  1){ 10}}
\put( 65,181){\line(  5,  2){ 10}}
\put( 75,185){\line(  5,  2){ 10}}
\put( 85,189){\line(  3,  1){ 10}}
\put( 95,192){\line(  4,  1){ 10}}
\put(105,194){\line(  5,  1){ 10}}
\put(115,196){\line(  6,  1){ 10}}
\put(125,198){\line(  1,  0){ 10}}
\put(135,198){\line(  6,  1){ 10}}
\put(145,200){\line(  1,  0){ 10}}
\put(155,200){\line(  6, -1){ 10}}
\put(165,198){\line(  1,  0){ 10}}
\put(175,198){\line(  6, -1){ 10}}
\put(185,196){\line(  5, -1){ 10}}
\put(195,194){\line(  4, -1){ 10}}
\put(205,192){\line(  3, -1){ 10}}
\put(215,189){\line(  5, -2){ 10}}
\put(225,185){\line(  5, -2){ 10}}
\put(235,181){\line(  2, -1){ 10}}
\put(245,176){\line(  5, -3){ 10}}
\put(255,170){\line(  3, -2){ 10}}
\put(265,163){\line(  4, -3){ 10}}
\put(275,155){\line(  6, -5){ 10}}
\put(285,147){\line(  1, -1){ 10}}
\put(295,137){\line(  5, -6){ 10}}
\put(305,125){\line(  4, -5){ 10}}
\put(315,113){\line(  3, -5){ 10}}
\put(325, 96){\line(  1, -2){ 10}}
\put(335, 76){\line(  1, -3){ 10}}

\put(-45, 55){\line(  2,  5){ 10}}
\put(-25,100){\line(  3,  5){ 10}}
\put( -5,130){\line(  5,  6){ 10}}
\put( 15,150){\line(  6,  5){ 10}}
\put( 35,166){\line(  3,  2){ 10}}
\put( 55,178){\line(  2,  1){ 10}}
\put( 75,188){\line(  3,  1){ 10}}
\put( 95,195){\line(  4,  1){ 10}}
\put(115,199){\line(  6,  1){ 10}}
\put(135,201){\line(  6,  1){ 10}}
\put(155,202){\line(  6, -1){ 10}}
\put(175,201){\line(  6, -1){ 10}}
\put(195,197){\line(  4, -1){ 10}}
\put(215,191){\line(  3, -1){ 10}}
\put(235,183){\line(  2, -1){ 10}}
\put(255,173){\line(  3, -2){ 10}}
\put(275,159){\line(  6, -5){ 10}}
\put(295,140){\line(  1, -1){ 10}}
\put(315,117){\line(  2, -3){ 10}}
\put(335, 82){\line(  2, -5){ 10}}


\end{picture}
}

{\Large \vspace*{5cm} }

\begin{center}
{\Large Fig.1 }
\end{center}

\newpage \vspace*{2cm}
\qquad\qquad {\Large 
\begin{picture}(300,200)

\thinlines
\put(-50,0){\line(1,0){400}}
\put(-50,200){\line(1,0){400}}
\put(-50,0){\line(0,1){200}}
\put(350,0){\line(0,1){200}}
\put(-30,0){\line(0,1){8}}
\put(-10,0){\line(0,1){8}}
\put(10,0){\line(0,1){8}}
\put(30,0){\line(0,1){8}}
\put(50,0){\line(0,1){8}}
\put(70,0){\line(0,1){8}}
\put(90,0){\line(0,1){8}}
\put(110,0){\line(0,1){8}}
\put(130,0){\line(0,1){8}}
\put(150,0){\line(0,1){8}}
\put(170,0){\line(0,1){8}}
\put(190,0){\line(0,1){8}}
\put(210,0){\line(0,1){8}}
\put(230,0){\line(0,1){8}}
\put(250,0){\line(0,1){8}}
\put(270,0){\line(0,1){8}}
\put(290,0){\line(0,1){8}}
\put(310,0){\line(0,1){8}}
\put(330,0){\line(0,1){8}}

\put(-50,40){\line(1,0){6}}
\put(-75,36){$0.2$}
\put(-50,80){\line(1,0){6}}
\put(-75,76){$0.4$}
\put(-50,120){\line(1,0){6}}
\put(-75,116){$0.6$}
\put(-50,160){\line(1,0){6}}
\put(-75,156){$0.8$}
\put(-75,196){$1.0$}
\put(-45,180){$g(x)$}

\put(315,15){$x$}
\put(147,-18){$0$}
\put(-60,-18){$-1.0$}
\put(340,-18){$1.0$}

\thicklines
\put(-45,  9){\line(  6,  5){ 10}}
\put(-35, 17){\line(  6,  5){ 10}}
\put(-25, 25){\line(  4,  3){ 10}}
\put(-15, 33){\line(  5,  4){ 10}}
\put( -5, 41){\line(  4,  3){ 10}}
\put(  5, 48){\line(  5,  4){ 10}}
\put( 15, 56){\line(  5,  4){ 10}}
\put( 25, 64){\line(  4,  3){ 10}}
\put( 35, 72){\line(  5,  4){ 10}}
\put( 45, 80){\line(  5,  4){ 10}}
\put( 55, 88){\line(  5,  4){ 10}}
\put( 65, 96){\line(  4,  3){ 10}}
\put( 75,103){\line(  4,  3){ 10}}
\put( 85,111){\line(  4,  3){ 10}}
\put( 95,118){\line(  4,  3){ 10}}
\put(105,126){\line(  3,  2){ 10}}
\put(115,133){\line(  4,  3){ 10}}
\put(125,140){\line(  5,  3){ 10}}
\put(135,146){\line(  3,  2){ 10}}
\put(145,153){\line(  5,  3){ 10}}
\put(155,159){\line(  2,  1){ 10}}
\put(165,164){\line(  2,  1){ 10}}
\put(175,169){\line(  2,  1){ 10}}
\put(185,174){\line(  3,  1){ 10}}
\put(195,177){\line(  3,  1){ 10}}
\put(205,180){\line(  4,  1){ 10}}
\put(215,183){\line(  5,  1){ 10}}
\put(225,185){\line(  6,  1){ 10}}
\put(235,187){\line(  6, -1){ 10}}
\put(245,185){\line(  1,  0){ 10}}
\put(255,185){\line(  4, -1){ 10}}
\put(265,182){\line(  5, -2){ 10}}
\put(275,178){\line(  5, -3){ 10}}
\put(285,172){\line(  3, -2){ 10}}
\put(295,166){\line(  1, -1){ 10}}
\put(305,156){\line(  3, -4){ 10}}
\put(315,142){\line(  3, -5){ 10}}
\put(325,126){\line(  2, -5){ 10}}
\put(335,101){\line(  1, -4){ 10}}

\put(-45, 10){\line(  4,  3){ 10}}
\put(-25, 25){\line(  4,  3){ 10}}
\put( -5, 40){\line(  4,  3){ 10}}
\put( 15, 55){\line(  5,  4){ 10}}
\put( 35, 71){\line(  5,  4){ 10}}
\put( 55, 87){\line(  5,  4){ 10}}
\put( 75,102){\line(  4,  3){ 10}}
\put( 95,118){\line(  4,  3){ 10}}
\put(115,132){\line(  3,  2){ 10}}
\put(135,145){\line(  3,  2){ 10}}
\put(155,158){\line(  5,  3){ 10}}
\put(175,169){\line(  5,  2){ 10}}
\put(195,177){\line(  5,  2){ 10}}
\put(215,183){\line(  5,  1){ 10}}
\put(235,187){\line(  1,  0){ 10}}
\put(255,185){\line(  6, -1){ 10}}
\put(275,180){\line(  2, -1){ 10}}
\put(295,169){\line(  1, -1){ 10}}
\put(315,147){\line(  3, -5){ 10}}
\put(330,120){\line(  2, -5){ 6}}
\put(338,100){\line(  1, -4){ 5}}


\end{picture}
}

{\Large \vspace*{5cm} }

\begin{center}
{\Large Fig.2 }
\end{center}

\newpage \vspace*{2cm}
\qquad\qquad {\Large 
\begin{picture}(300,200)

\thinlines
\put(-50,0){\line(1,0){400}}
\put(-50,240){\line(1,0){400}}
\put(-50,0){\line(0,1){240}}
\put(350,0){\line(0,1){240}}
\put(-50,0){\line(0,-1){200}}
\put(350,0){\line(0,-1){200}}
\put(-50,-200){\line(1,0){400}}

\put(-30,0){\line(0,1){8}}
\put(-10,0){\line(0,1){8}}
\put(10,0){\line(0,1){8}}
\put(30,0){\line(0,1){8}}
\put(50,0){\line(0,1){8}}
\put(70,0){\line(0,1){8}}
\put(90,0){\line(0,1){8}}
\put(110,0){\line(0,1){8}}
\put(130,0){\line(0,1){8}}
\put(150,0){\line(0,1){8}}
\put(170,0){\line(0,1){8}}
\put(190,0){\line(0,1){8}}
\put(210,0){\line(0,1){8}}
\put(230,0){\line(0,1){8}}
\put(250,0){\line(0,1){8}}
\put(270,0){\line(0,1){8}}
\put(290,0){\line(0,1){8}}
\put(310,0){\line(0,1){8}}
\put(330,0){\line(0,1){8}}

\put(-50,40){\line(1,0){6}}
\put(-75,36){$0.2$}
\put(-50,80){\line(1,0){6}}
\put(-75,76){$0.4$}
\put(-50,120){\line(1,0){6}}
\put(-75,116){$0.6$}
\put(-50,160){\line(1,0){6}}
\put(-75,156){$0.8$}
\put(-50,200){\line(1,0){6}}
\put(-75,196){$1.0$}
\put(-45,220){$Re[g(x)]$}
\put(-75,236){$1.2$}

\put(-50,-40){\line(1,0){6}}
\put(-92,-44){$-0.2$}
\put(-50,-80){\line(1,0){6}}
\put(-92,-84){$-0.4$}
\put(-50,-120){\line(1,0){6}}
\put(-92,-124){$-0.6$}
\put(-50,-160){\line(1,0){6}}
\put(-92,-164){$-0.8$}
\put(-92,-204){$-1.0$}
\put(-45,-180){$Im[g(x)]$}

\put(355,-2){$x$}
\put(147,-18){$0$}

\thicklines
\put(-45, 62){\line(  1,  3){ 10}}
\put(-35, 92){\line(  1,  2){ 10}}
\put(-25,112){\line(  1,  2){ 10}}
\put(-15,132){\line(  2,  3){ 10}}
\put( -5,147){\line(  5,  6){ 10}}
\put(  5,159){\line(  5,  6){ 10}}
\put( 15,171){\line(  6,  5){ 10}}
\put( 25,180){\line(  1,  1){ 10}}
\put( 35,190){\line(  3,  2){ 10}}
\put( 45,196){\line(  3,  2){ 10}}
\put( 55,203){\line(  5,  3){ 10}}
\put( 65,209){\line(  2,  1){ 10}}
\put( 75,214){\line(  5,  2){ 10}}
\put( 85,218){\line(  5,  2){ 10}}
\put( 95,222){\line(  3,  1){ 10}}
\put(105,225){\line(  5,  1){ 10}}
\put(115,227){\line(  5,  1){ 10}}
\put(125,229){\line(  6,  1){ 10}}
\put(135,231){\line(  1,  0){ 10}}
\put(145,231){\line(  1,  0){ 10}}
\put(155,231){\line(  1,  0){ 10}}
\put(165,231){\line(  6, -1){ 10}}
\put(175,229){\line(  5, -1){ 10}}
\put(185,227){\line(  4, -1){ 10}}
\put(195,225){\line(  4, -1){ 10}}
\put(205,222){\line(  5, -2){ 10}}
\put(215,218){\line(  5, -2){ 10}}
\put(225,214){\line(  2, -1){ 10}}
\put(235,209){\line(  5, -3){ 10}}
\put(245,203){\line(  3, -2){ 10}}
\put(255,197){\line(  5, -4){ 10}}
\put(265,189){\line(  6, -5){ 10}}
\put(275,180){\line(  1, -1){ 10}}
\put(285,170){\line(  1, -1){ 10}}
\put(295,160){\line(  3, -4){ 10}}
\put(305,147){\line(  2, -3){ 10}}
\put(315,132){\line(  3, -5){ 10}}
\put(325,115){\line(  2, -5){ 10}}
\put(335, 90){\line(  1, -3){ 10}}

\put(-45, -6){\line(  3, -1){ 10}}
\put(-35,-10){\line(  4, -1){ 10}}
\put(-25,-12){\line(  5, -1){ 10}}
\put(-15,-14){\line(  6, -1){ 10}}
\put( -5,-16){\line(  6, -1){ 10}}
\put(  5,-18){\line(  6, -1){ 10}}
\put( 15,-19){\line(  1,  0){ 10}}
\put( 25,-19){\line(  6, -1){ 10}}
\put( 35,-21){\line(  1,  0){ 10}}
\put( 45,-21){\line(  6, -1){ 10}}
\put( 55,-23){\line(  1,  0){ 10}}
\put( 65,-23){\line(  1,  0){ 10}}
\put( 75,-23){\line(  6, -1){ 10}}
\put( 85,-24){\line(  1,  0){ 10}}
\put( 95,-24){\line(  1,  0){ 10}}
\put(105,-24){\line(  1,  0){ 10}}
\put(115,-24){\line(  1,  0){ 10}}
\put(125,-24){\line(  1,  0){ 10}}
\put(135,-24){\line(  6, -1){ 10}}
\put(145,-26){\line(  1,  0){ 10}}
\put(155,-26){\line(  6,  1){ 10}}
\put(165,-24){\line(  1,  0){ 10}}
\put(175,-24){\line(  1,  0){ 10}}
\put(185,-24){\line(  1,  0){ 10}}
\put(195,-24){\line(  1,  0){ 10}}
\put(205,-24){\line(  1,  0){ 10}}
\put(215,-24){\line(  6,  1){ 10}}
\put(225,-23){\line(  1,  0){ 10}}
\put(235,-23){\line(  1,  0){ 10}}
\put(245,-23){\line(  6,  1){ 10}}
\put(255,-21){\line(  1,  0){ 10}}
\put(265,-21){\line(  6,  1){ 10}}
\put(275,-19){\line(  6,  1){ 10}}
\put(285,-18){\line(  1,  0){ 10}}
\put(295,-18){\line(  6,  1){ 10}}
\put(305,-16){\line(  6,  1){ 10}}
\put(315,-14){\line(  5,  1){ 10}}
\put(325,-12){\line(  4,  1){ 10}}
\put(335,-10){\line(  3,  1){ 10}}

\put(-45, 30){\line(  2,  3){ 10}}
\put(-35, 45){\line(  5,  6){ 10}}
\put(-25, 57){\line(  6,  5){ 10}}
\put(-15, 65){\line(  5,  4){ 10}}
\put( -5, 73){\line(  3,  2){ 10}}
\put(  5, 80){\line(  5,  3){ 10}}
\put( 15, 86){\line(  2,  1){ 10}}
\put( 25, 91){\line(  2,  1){ 10}}
\put( 35, 96){\line(  5,  2){ 10}}
\put( 45,100){\line(  5,  2){ 10}}
\put( 55,104){\line(  3,  1){ 10}}
\put( 65,107){\line(  3,  1){ 10}}
\put( 75,111){\line(  4,  1){ 10}}
\put( 85,113){\line(  5,  1){ 10}}
\put( 95,115){\line(  5,  1){ 10}}
\put(105,117){\line(  6,  1){ 10}}
\put(115,119){\line(  1,  0){ 10}}
\put(125,119){\line(  6,  1){ 10}}
\put(135,121){\line(  1,  0){ 10}}
\put(145,121){\line(  1,  0){ 10}}
\put(155,121){\line(  1,  0){ 10}}
\put(165,121){\line(  6, -1){ 10}}
\put(175,119){\line(  1,  0){ 10}}
\put(185,119){\line(  6, -1){ 10}}
\put(195,117){\line(  5, -1){ 10}}
\put(205,115){\line(  5, -1){ 10}}
\put(215,113){\line(  4, -1){ 10}}
\put(225,111){\line(  3, -1){ 10}}
\put(235,107){\line(  3, -1){ 10}}
\put(245,104){\line(  3, -1){ 10}}
\put(255,101){\line(  2, -1){ 10}}
\put(265, 96){\line(  5, -2){ 10}}
\put(275, 92){\line(  5, -3){ 10}}
\put(285, 86){\line(  5, -3){ 10}}
\put(295, 80){\line(  3, -2){ 10}}
\put(305, 73){\line(  4, -3){ 10}}
\put(315, 66){\line(  1, -1){ 10}}
\put(325, 56){\line(  1, -1){ 10}}
\put(335, 46){\line(  2, -3){ 10}}

\put(-45, -43){\line(  2, -5){ 10}}
\put(-35, -68){\line(  2, -3){ 10}}
\put(-25, -83){\line(  2, -3){ 10}}
\put(-15, -98){\line(  4, -5){ 10}}
\put( -5,-111){\line(  5, -6){ 10}}
\put(  5,-123){\line(  1, -1){ 10}}
\put( 15,-133){\line(  1, -1){ 10}}
\put( 25,-143){\line(  6, -5){ 10}}
\put( 35,-151){\line(  6, -5){ 10}}
\put( 45,-159){\line(  4, -3){ 10}}
\put( 55,-167){\line(  3, -2){ 10}}
\put( 65,-174){\line(  5, -3){ 10}}
\put( 75,-180){\line(  2, -1){ 10}}
\put( 85,-185){\line(  5, -2){ 10}}
\put( 95,-189){\line(  5, -2){ 10}}
\put(105,-193){\line(  4, -1){ 10}}
\put(115,-195){\line(  4, -1){ 10}}
\put(125,-198){\line(  6, -1){ 10}}
\put(135,-199){\line(  1,  0){ 10}}
\put(145,-199){\line(  1,  0){ 10}}
\put(155,-199){\line(  1,  0){ 10}}
\put(165,-199){\line(  6,  1){ 10}}
\put(175,-198){\line(  5,  1){ 10}}
\put(185,-196){\line(  3,  1){ 10}}
\put(195,-192){\line(  3,  1){ 10}}
\put(205,-189){\line(  5,  2){ 10}}
\put(215,-185){\line(  5,  3){ 10}}
\put(225,-179){\line(  2,  1){ 10}}
\put(235,-174){\line(  3,  2){ 10}}
\put(245,-167){\line(  4,  3){ 10}}
\put(255,-160){\line(  5,  4){ 10}}
\put(265,-152){\line(  6,  5){ 10}}
\put(275,-143){\line(  1,  1){ 10}}
\put(285,-133){\line(  1,  1){ 10}}
\put(295,-123){\line(  4,  5){ 10}}
\put(305,-111){\line(  3,  4){ 10}}
\put(315, -98){\line(  2,  3){ 10}}
\put(325, -83){\line(  3,  5){ 10}}
\put(335, -66){\line(  2,  5){ 10}}

\put(-45,  12){\line(  2,  1){ 10}}
\put(-35,  17){\line(  3,  1){ 10}}
\put(-25,  21){\line(  6,  1){ 10}}
\put(-15,  22){\line(  1,  0){ 10}}
\put( -5,  22){\line(  1,  0){ 10}}
\put(  5,  22){\line(  1,  0){ 10}}
\put( 15,  22){\line(  1,  0){ 10}}
\put( 25,  22){\line(  6, -1){ 10}}
\put( 35,  21){\line(  1,  0){ 10}}
\put( 45,  21){\line(  6, -1){ 10}}
\put( 55,  19){\line(  6, -1){ 10}}
\put( 65,  17){\line(  1,  0){ 10}}
\put( 75,  17){\line(  6, -1){ 10}}
\put( 85,  16){\line(  6, -1){ 10}}
\put( 95,  14){\line(  1,  0){ 10}}
\put(105,  14){\line(  6, -1){ 10}}
\put(115,  12){\line(  1,  0){ 10}}
\put(125,  12){\line(  1,  0){ 10}}
\put(135,  12){\line(  1,  0){ 10}}
\put(145,  12){\line(  1,  0){ 10}}
\put(155,  12){\line(  1,  0){ 10}}
\put(165,  12){\line(  1,  0){ 10}}
\put(175,  12){\line(  1,  0){ 10}}
\put(185,  12){\line(  6,  1){ 10}}
\put(195,  14){\line(  1,  0){ 10}}
\put(205,  14){\line(  6,  1){ 10}}
\put(215,  16){\line(  6,  1){ 10}}
\put(225,  17){\line(  1,  0){ 10}}
\put(235,  17){\line(  6,  1){ 10}}
\put(245,  19){\line(  6,  1){ 10}}
\put(255,  21){\line(  1,  0){ 10}}
\put(265,  21){\line(  6,  1){ 10}}
\put(275,  22){\line(  1,  0){ 10}}
\put(285,  22){\line(  1,  0){ 10}}
\put(295,  22){\line(  1,  0){ 10}}
\put(305,  22){\line(  1,  0){ 10}}
\put(315,  22){\line(  5, -1){ 10}}
\put(325,  20){\line(  3, -1){ 10}}
\put(335,  17){\line(  2, -1){ 10}}

\put(-45, -18){\line(  1, -1){ 10}}
\put(-35, -28){\line(  1, -1){ 10}}
\put(-25, -38){\line(  6, -5){ 10}}
\put(-15, -47){\line(  6, -5){ 10}}
\put( -5, -55){\line(  5, -4){ 10}}
\put(  5, -63){\line(  4, -3){ 10}}
\put( 15, -71){\line(  3, -2){ 10}}
\put( 25, -77){\line(  4, -3){ 10}}
\put( 35, -85){\line(  5, -3){ 10}}
\put( 45, -91){\line(  5, -3){ 10}}
\put( 55, -97){\line(  2, -1){ 10}}
\put( 65,-102){\line(  5, -3){ 10}}
\put( 75,-108){\line(  5, -2){ 10}}
\put( 85,-112){\line(  5, -2){ 10}}
\put( 95,-116){\line(  3, -1){ 10}}
\put(105,-119){\line(  4, -1){ 10}}
\put(115,-122){\line(  5, -1){ 10}}
\put(125,-124){\line(  6, -1){ 10}}
\put(135,-125){\line(  1,  0){ 10}}
\put(145,-125){\line(  1,  0){ 10}}
\put(155,-125){\line(  1,  0){ 10}}
\put(165,-125){\line(  6,  1){ 10}}
\put(175,-124){\line(  5,  1){ 10}}
\put(185,-122){\line(  4,  1){ 10}}
\put(195,-119){\line(  3,  1){ 10}}
\put(205,-116){\line(  5,  2){ 10}}
\put(215,-112){\line(  5,  2){ 10}}
\put(225,-108){\line(  5,  3){ 10}}
\put(235,-102){\line(  2,  1){ 10}}
\put(245, -97){\line(  5,  3){ 10}}
\put(255, -91){\line(  3,  2){ 10}}
\put(265, -84){\line(  3,  2){ 10}}
\put(275, -77){\line(  3,  2){ 10}}
\put(285, -71){\line(  4,  3){ 10}}
\put(295, -63){\line(  5,  4){ 10}}
\put(305, -55){\line(  5,  4){ 10}}
\put(315, -47){\line(  6,  5){ 10}}
\put(325, -39){\line(  1,  1){ 10}}
\put(335, -29){\line(  1,  1){ 10}}

\put(263,100){$2$}
\put(280,-155){$2$}
\put(290,170){$1$}
\put(197,-40){$1$}
\put(230,-120){$3$}
\put(250,25){$3$}

\end{picture}
}

{\Large \vspace*{8cm} }

\begin{center}
{\Large Fig.3 }
\end{center}

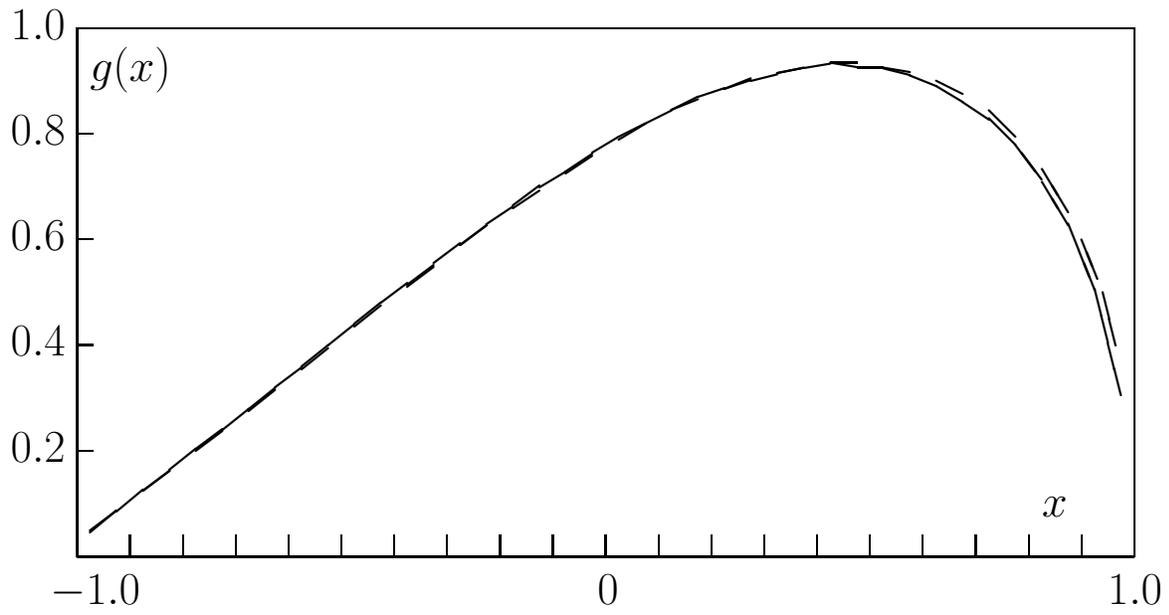
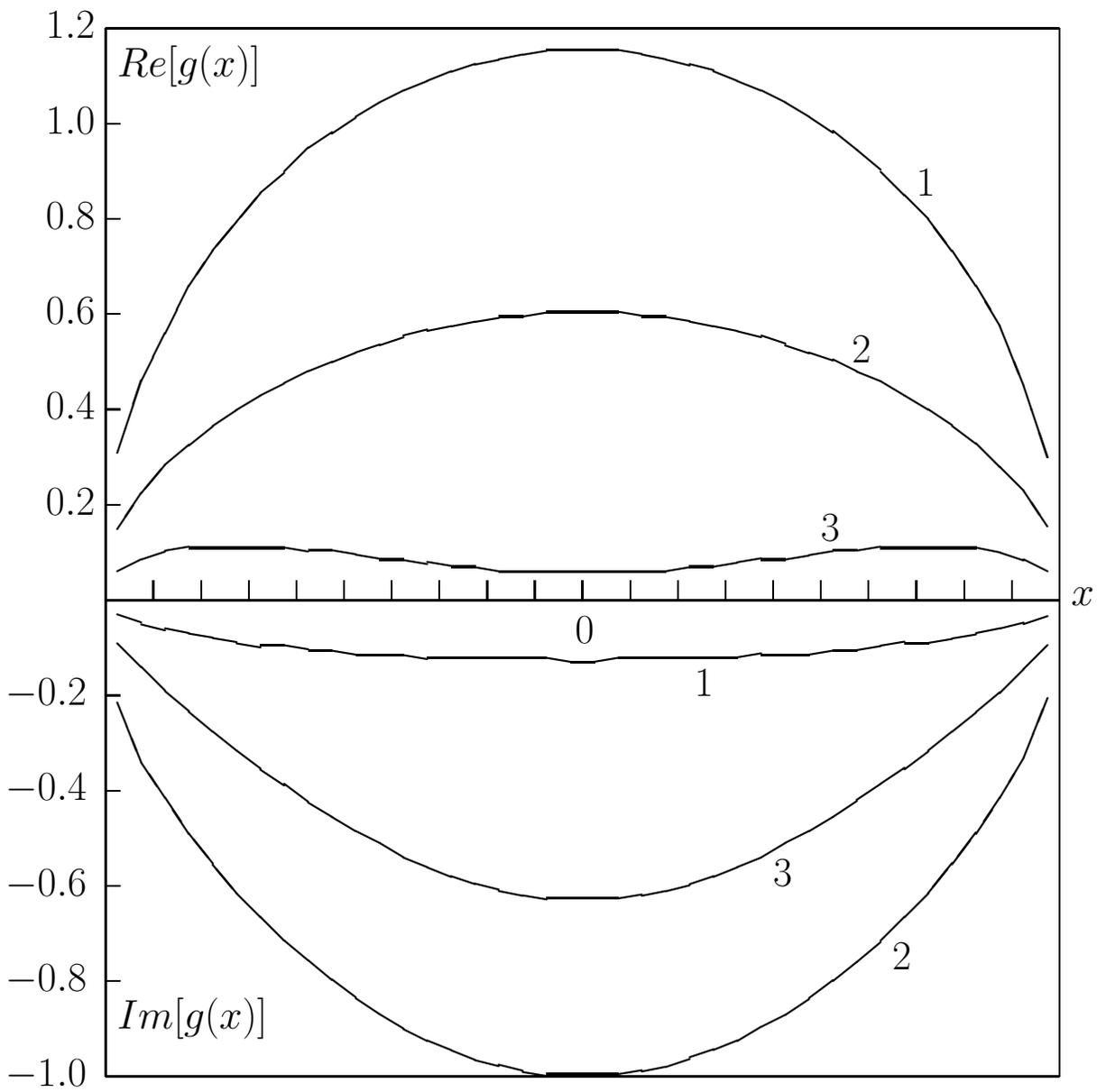
\end{document}